\documentclass[12pt]{iopart}

\usepackage{graphicx}
\usepackage{bm}
\usepackage{amssymb}

\begin{document}
\title[Ambient temperature fluctuations exhibit inverse-cubic spectral behavior]{Ambient temperature fluctuations exhibit inverse-cubic spectral behavior over time scales from a minute to a day\footnote{\emph{J. Phys. Commun.} 4 (2020) 041001; Published 3 April; https://doi.org/10.1088/2399-6528/ab82b7.\\ Original content from this work may be used under the terms of the Creative Commons Attribution 4.0 license. Any further distribution of this work must maintain attribution to the author(s) and the title of the work, journal citation and DOI.}}
\author{Steven B Lowen$^{1,2,3}$, Nishant Mohan$^{1,4,5}$
%\footnote{Present address: Wasatch Photonics, Morrisville, North Carolina 27560, United States of America} 
and Malvin Carl Teich$^{1,2,4,6,7}$}
\address{$^1$ Boston University, Photonics Center, Boston, Massachusetts 02215, United States of America}
\address{$^2$ Boston University, Department of Electrical \& Computer Engineering, Boston, Massachusetts 02215, United States of America}
\address{$^3$ Harvard Medical School, Boston, Massachusetts 02115, United States of America}
\address{$^4$ Boston University, Department of Biomedical Engineering, Boston, Massachusetts 02215, United States of America}
\address{$^5$ Massachusetts General Hospital, Wellman Center for Photomedicine, Boston, Massachusetts 02114, United States of America}
\address{$^6$ Boston University, Department of Physics, Boston, Massachusetts 02215, United States of America}
\address{$^7$ Columbia University, Department of Electrical Engineering, New York, New York 10027, United States of America}
\ead{teich@bu.edu}

\vspace{5mm}

\vspace{6mm}
\begin{abstract}
Ambient temperature fluctuations are of importance in a wide variety of scientific and technological arenas. In a series of experiments carried out in our laboratory over an 18-month period, we discovered that these fluctuations exhibit $1/f^3$ spectral behavior over the frequency range $1.0 \times 10^{-5} \le f \le 2.5 \times 10^{-2}$~Hz, corresponding to $40$~s $ \le T_f \le 1.2$~d, where $T_f = 1/f$. This result emerges over a broad range of conditions. For longer time periods, $1.2 < T_f \le 11.6$~d, corresponding to the frequency range $1.0 \times 10^{-6} \le f < 1.0 \times 10^{-5}$ Hz, we observed $1/f^2$ spectral behavior. This latter result is in agreement with that observed in data collected at European weather stations. Scalograms computed from our data are consistent with the periodograms.
\end{abstract}

\maketitle

\section{Introduction}
Ambient temperature plays an important role in governing the behavior of many biological and physical phenomena. It has a strong impact on a vast array of processes, in areas as diverse as neuronal signaling \cite{hodgkin49}, vaccine production \cite{josefsberg12},
semiconductor physics \cite{saleh19}, and gravitational-wave detection in the presence of thermal noise \cite{abbott16}. In the course of conducting a series of long-duration photon-counting experiments that involved semiconductor optics, it was essential to characterize the ambient temperature fluctuations in the local environment.
The low-frequency spectra of these temperature fluctuations, which to our knowledge have not been previously investigated, were found to exhibit a universal power-law form.

In section~\ref{sec:methods} we discuss the methods used to collect and analyze the data. In section~\ref{sec:results} we present the results, which take the form of periodograms and wavelet-based scalograms. The conclusion and discussion are provided in section~\ref{sec:concl}.

\section{Methods} \label{sec:methods}

\subsection{Laboratory and experimental configuration}
The experiments were conducted in a standard-issue laboratory space on the 7th floor of the Photonics Center at Boston University.  The laboratory was a closed-plan, 75-m$^2$ facility that contained an optical bench located near its center, on which all experiments were carried out. Standard photonic and electronic instrumentation, including light sources, photodetectors, power supplies, and computers, were placed on the optical bench or on shelving above it, or on a separate bench to its side.  Equipment pilot lights were extinguished or covered. The temperature sensors were located on the optical bench, in the open. 

The laboratory space was dedicated exclusively to this project during its 18-month duration. Each individual temperature experiment took place over an 11.6-day uninterrupted period, during which janitorial services were suspended and all personnel (including the experimenters) were excluded from the laboratory. Experiments were conducted both when the HVAC (`bang-bang' heating, ventilation, and air conditioning) system was engaged to provide the laboratory with heating or cooling, as well as when the HVAC system was disabled. Analysis revealed that the temperature fluctuations collected under all three conditions (HVAC in heating mode, HVAC in cooling mode, and HVAC disabled) were indistinguishable. Similarly, the temperature fluctuations exhibited no discernible dependence on the season when the data were collected (fall, winter, spring, summer). The observed intrinsic temperature fluctuations were evidently decoupled from external temperature variations. The temperature measurements were conducted concomitantly with a series of photon-counting experiments that involved semiconductor optics, the results of which are reported separately \cite{mohan2020}.  

\subsection{Temperature measurement systems} \label{ssec:measurement}
Two independent temperature-measurement systems were employed:

\subsubsection{Thorlabs integrated-circuit resistance temperature detector.}

In this system, the temperature was measured with a Thorlabs (Newton, New Jersey) Integrated-Circuit Thin-Film-Resistor Temperature Transducer (Model AD590), which provides an output current proportional to absolute temperature. Used in conjunction with a Thorlabs Laser Diode and Temperature Controller (Model ITC502) operated in an open-loop configuration, this system has a temperature resolution of $0.01^\circ$C, repeatability of $\pm0.1^\circ$C, long-term drift of $\pm0.1^\circ$C, and an accuracy of $\pm3.0^\circ$C. Its operating range stretches from $-45^\circ$C to $+145^\circ$C. Temperature readings were recorded every $5$~s and transferred to a laboratory computer via a National Instruments GPIB (General Purpose Interface Bus) \cite{mohan10}.

\subsubsection{ThermoWorks resistance temperature detector.}

Independent temperature measurements were made throughout the duration of each experiment using a ThermoWorks (Lindon, Utah) Temperature Data Logger (Model TEMP101A) that uses a resistance temperature detector. This device has a temperature resolution of $0.01^\circ$C and an accuracy of $\pm0.5^\circ$C. Its operating range extends from $-40^\circ$C to $+80^\circ$C and it has a storage capacity of $10^6$ readings. Temperature readings were again made every 5~s; in this case they were transferred to the laboratory computer in real time via a USB connector cable. Of nearly $3$~million ThermoWorks temperature readings, $26$ displayed a spurious value of $639.25^\circ$C, which randomly occurred over several files. These readings were replaced with the previous non-spurious readings.

\subsection{Computational procedures}

\subsubsection{Periodograms.}

Periodograms represent estimates of the power spectral density versus frequency \cite{oppenheim09}. Ambient temperature periodograms for 13 mean-temperature data sets collected in our laboratory using the temperature measurement systems described in section~\ref{ssec:measurement} are reported in section~\ref{ssec:periodogram}. The data were not truncated to avoid the introduction of bias. The periodograms were computed from the raw data via a procedure similar to that described in section~12.3.9 of Lowen \& Teich \cite{lowen05}, a summary of which is provided below. The data used to construct the periodograms were deliberately not detrended to avoid inadvertently diluting or eliminating the fluctuations corresponding to the power-law behavior while attempting to remove nonstationarities, as explained in section~2.8.4 of Lowen \& Teich \cite{lowen05}. The periodograms were not normalized.

\begin{enumerate}
  \item The data were demeaned.
  \item The data were multiplied by a Hann (raised cosine) window so that the values at the edges were reduced to zero; this extended the frequency exponent that could be accommodated from $-2$ to $-6$, which comfortably includes $-3$, as explained in Secs.~12.3.9 and A.8.1 of Lowen \& Teich \cite{lowen05}.
  \item The data were zero-padded to the next largest power of two.
  \item The fast-Fourier transform (FFT) was computed.
  \item The square of the absolute magnitude was calculated.
  \item The resulting values were divided by the number of points in the
      original data set.
  \item Values with abscissas within a factor of 1.02 were averaged to yield a single periodogram abscissa and ordinate.
\end{enumerate}

\subsubsection{Scalograms.}

Daubechies 4-tap scalograms versus scale for the same 13 mean-temperature data sets are reported in section~\ref{ssec:scalogram}. The scalograms were computed from the raw data via a procedure similar to that described in section~5.4.4 of Lowen \& Teich \cite{lowen05}, a summary of which is provided below. As with the periodograms, neither detrending nor normalization was used in constructing the scalograms.

\begin{enumerate}
    \item A Daubechies 4-tap wavelet transform \cite{daubechies92} was carried out, using all available scales. 
    \item All elements of the wavelet transform that spanned the edges of the data set were discarded.
    \item The square of all remaining elements were calculated. Since wavelet transforms are zero-mean by construction \cite{haar10,graps95,ryan19}, this allowed the wavelet variance to be directly determined.  
    \item All such elements with the same scale were averaged.
\end{enumerate} 

\subsection{Notation} \label{ssec:notation}

\begin{enumerate}
  \item Temperature is denoted $\mathcal{T}$ and the temperature periodogram is denoted $S_{\mathcal{T}}(f)$, where $f$ is the periodogram harmonic frequency.
  \item The inverse of the periodogram frequency $f$ is denoted $T_f$. Diurnal temperature fluctuations, which correspond to $T_f = 1$~d = $86\,400$~s, thus appear in the periodogram at $f = 1/T_f = 1.16 \times 10^{-5}$~Hz, as discussed in section~\ref{ssec:periodogram}.
  \item Sampling that takes place every $T_s$~s results in a maximum allowed frequency $\frac12 /T_s$, as established by the Nyquist limit \cite{oppenheim09}. In section~\ref{ssec:periodogram} we have $T_s = 5$~s so the highest allowed frequency is $\frac12 / (5~\mathrm{s}) = 0.10$~Hz, whereas in section~\ref{sec:concl} we have $T_s = 1$~d so the highest allowed frequency is $\frac12 / (86400~\mathrm{s}) = 5.79 \times 10^{-6}$~Hz.
  \item The Daubechies 4-tap wavelet variance is denoted $A_{\mathcal{T}}(T)$, where $T$ is the wavelet scale.
\end{enumerate}

\section{Results} \label{sec:results}

\subsection{Ambient Temperature Periodograms} \label{ssec:periodogram}
\begin{figure}[ht]
  \centering
  \includegraphics[width=\linewidth]{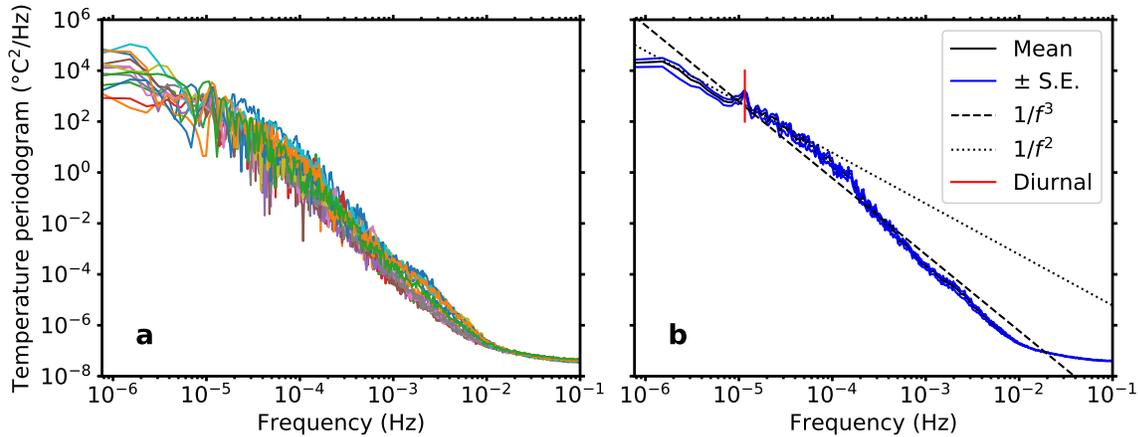}
  \caption{Temperature periodograms plotted on doubly-logarithmic coordinates.
  (a)~Unaveraged periodograms $S_{\mathcal{T}}(f)$ versus frequency $f$, using the Thorlabs temperature detector. Though measured under different laboratory conditions, all 13 plots closely follow each other. (b)~Averaged periodogram $\overline{S}_{\mathcal{T}}(f)$ versus frequency $f$. The solid black curve is the mean of all 13 periodograms displayed in (a). The two blue curves hugging the black curve represent standard-error values ($\pm$~S.E.). The averaged periodogram decreases approximately as $1/f^2$ (dotted curve) over the frequency range $1.0 \times 10^{-6} \le f < 1.0 \times 10^{-5}$~Hz and approximately as $1/f^3$ (dashed curve) over the range $1.0 \times 10^{-5} \le f \le 2.5 \times 10^{-2}$~Hz. The vertical red line segment marks a small hump at $1.16 \times 10^{-5}$~Hz that corresponds to diurnal temperature variations (1~d = 86\,400~s).}   
  \label{bothtemppg}
  \end{figure}

The ambient temperature periodograms portrayed in figure~\ref{bothtemppg}(a), plotted one atop the other, represent $13$ data sets recorded under a variety of conditions. Collected with the Thorlabs temperature detector, all of the curves closely follow each other.
  
Indeed, the 13 curves presented in figure~\ref{bothtemppg}(a) are sufficiently similar that they may be averaged to improve the statistical accuracy of the data. To construct the averaged periodogram, a threshold of 198\,048 measurements was determined to be the best compromise between including as many data sets, and as much of each, as possible. At 5~s per measurement, that corresponds to $L = 990\,240$~s. 

The next largest power of 2 greater than 198\,048 is 262\,144, or 1\,310\,720~s. The lowest observable frequency is very close to 1~$\mu$Hz; it  could be reduced by zero-padding the data set. The highest observable frequency is the Nyquist limit $\frac12 / (5~\mathrm{s}) = 0.10$~Hz, as discussed in section~\ref{ssec:notation}. However, because of averaging of the values whose abscissas lie within a factor of 1.02, the highest displayed frequency is actually a bit smaller than 0.10~Hz, namely 0.098~Hz. 

The outcome of the averaging process, displayed in figure~\ref{bothtemppg}(b), exhibits one log-unit of $1/f^2$ behavior (dotted curve) over the frequency range $1.0 \times 10^{-6} \le f < 1.0 \times 10^{-5}$~Hz, corresponding to $1.2 < T_f \le 11.6$~d; and several log-units of roughly $1/f^3$ behavior (dashed curve) that stretches from $1.0 \times 10^{-5}$ to $2.5 \times 10^{-2}$~Hz, corresponding to $40$~s $\le T_f \le 1.2$~d. The small hump in the curve at $1.16 \times 10^{-5}$~Hz, designated by the vertical red line segment in the figure, corresponds to diurnal variations (1~d = 86\,400~s). Since the periodogram displayed in figure~\ref{bothtemppg}b becomes essentially flat for $f \gtrsim 2.5 \times 10^{-2}$~Hz, corresponding to $T_f \lesssim 40$~s, sampling the temperature at intervals shorter than $T_s = \frac12 T_f = 20$~s does not provide additional information.

The observation of temperature fluctuations with a $1/f^3$ spectral signature is unexpected and surprising, and to our knowledge has not been reported previously.

\subsection{Joint Temperature Measurements}\label{ssec:joint}
\begin{figure}[ht]
\centering
\includegraphics[width=\linewidth]{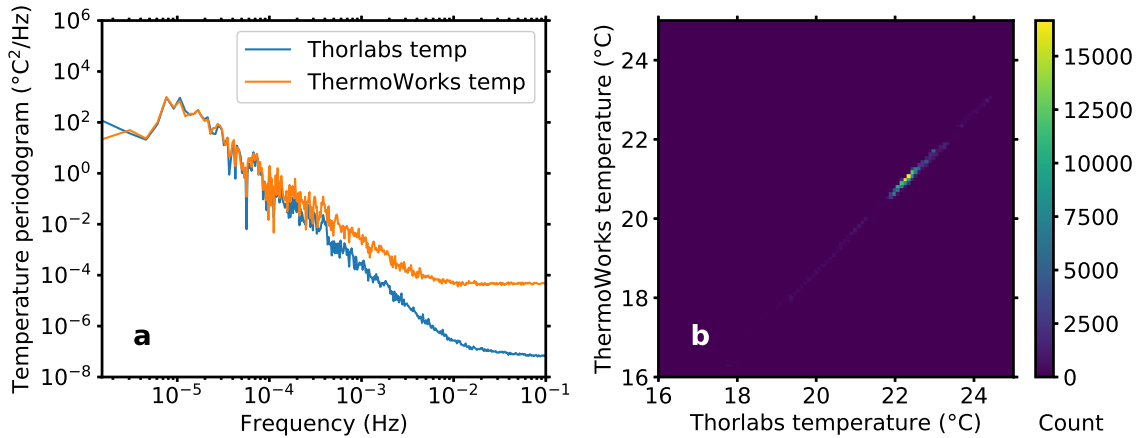}
\caption{Measurements carried out with two independent temperature-measurement systems. The data presented in this figure were collected in an experiment of duration 7.51~d. (a)~Temperature periodograms obtained with the Thorlabs and ThermoWorks temperature detectors, as indicated, plotted on doubly-logarithmic coordinates. The ThermoWorks device has higher baseline noise than the Thorlabs device. (b)~Joint temperature measurements. The temperatures registered by the Thorlabs and ThermoWorks devices, plotted on the abscissa and ordinate, respectively, are highly correlated.}
\label{both20111214_temp_pg}
\end{figure}

We now proceed to demonstrate that the two independent temperature-measurement systems, the Thorlabs and ThermoWorks resistance temperature detectors discussed in section~\ref{ssec:measurement}, yield similar results. The temperature periodograms displayed in figure~\ref{both20111214_temp_pg}(a) for both of these devices behave nearly indistinguishably for $f \lesssim 5 \times 10^{-4}$~Hz, the frequency at which the baseline noise of the ThermoWorks detector becomes appreciable. The Thorlabs detector registers fluctuations over nearly two additional log-units of frequency, as is clear from figure~\ref{bothtemppg}(b). 

 Joint temperature measurements for the two thermometric systems are displayed in figure~\ref{both20111214_temp_pg}(b). 
 The overall mean temperatures $\overline{\mathcal{T}}$ and coefficients of variation CV registered by the temperature detectors were $\overline{\mathcal{T}}_\mathrm{Thor} = 22.41^\circ$C %22.406
 and CV$_\mathrm{Thor} = 0.01262$ %0.0126241 
 for the Thorlabs device, and $\overline{\mathcal{T}}_\mathrm{Thermo} = 21.09^\circ$C %21.087 
 and CV$_\mathrm{Thermo} = 0.01257$ % 0.0125668
 for the ThermoWorks device. The measured overall mean temperatures differ by $1.32^\circ$C, which is within the margins of accuracy specified by the manufacturers: $\pm3.0^\circ$C for the Thorlabs device and $\pm0.5^\circ$C for the ThermoWorks device. The correlation coefficient for the  temperature measurements is $\rho = 0.9976$, %0.9976284522263859, 
 verifying that our measurement methods are sound and that our results are consistent. 

\subsection{Ambient Temperature Scalograms}\label{ssec:scalogram}

\begin{figure}[ht]
\centering
\includegraphics[width=\linewidth]{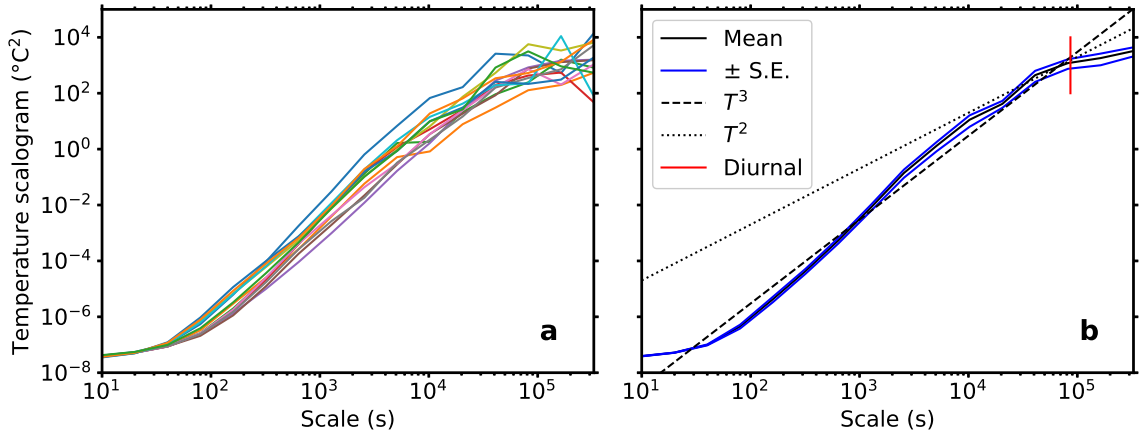}
\caption{Temperature scalograms plotted on doubly-logarithmic coordinates. (a)~Unaveraged Daubechies 4-tap wavelet variance $A_\mathcal{T}(T)$ versus scale $T$ using the Thorlabs temperature detector. As with the periodograms displayed in figure~\ref{bothtemppg}(a), all 13 plots closely follow each other even though they are measured under different laboratory conditions. (b)~Averaged Daubechies 4-tap wavelet variance $\overline{A}_\mathcal{T}(T)$ versus scale $T$. The solid black curve is the mean of all 13 scalograms displayed in (a). The two blue curves hugging the black curve represent standard-error values ($\pm$~S.E.). The averaged scalogram increases approximately as $T^3$ (dashed curve) over the range $24 \le T \le 6.7 \times 10^{4}$~s, in correspondence with the $1/f^3$ decrease of the averaged periodogram portrayed in figure~\ref{bothtemppg}(b). The vertical red line segment depicts the scale for diurnal temperature variations (1~d = 86\,400~s).}
\label{bothtempwv}
\end{figure}

The ambient temperature scalograms portrayed in figure~\ref{bothtempwv}(a), plotted one atop the other, represent the 13 data sets whose periodograms are presented in figure~\ref{bothtemppg}(a). The choice of the Daubechies 4-tap wavelet variance for the scalogram is governed by a number of considerations, including the extended power-law exponent it accommodates and its insensitivity to constant values and linear trends \cite{vetterli95,teich95}. 

As with the periodograms, all of the scalograms in figure~\ref{bothtempwv}(a) follow each other sufficiently closely that they can be averaged to improve the statistical accuracy of the data. The outcome of the averaging process, displayed in figure~\ref{bothtempwv}(b), yields a result that grows roughly as $T^3$ over several log-units (dashed curve), stretching from $T = 24$~s to $T = 6.7 \times 10^{4}$~s. The maximum permitted scale, corresponding to 1 sample, is 327\,680~s or 3.8~d. 

The observed power-law growth as $A(T) \propto T^3$ lies well below growth as $T^5$, where 5 is the maximum exponent permitted for the increase of the Daubechies 4-tap wavelet variance with scale and signals nonstationary behavior, as discussed in section~5.2.5 of Lowen \& Teich \cite{lowen05}. The cubic form of the observed scalogram corresponds to the inverse-cubic form of the observed periodogram. 

The averaged wavelet variance $\overline{A}_\mathcal{T}(T)$ displayed in figure~\ref{bothtempwv}(b) exhibits a small-scale cutoff at $T_A \approx 24$~s that separates behavior as $T^0$ from growth as $T^3$, and a large-scale cutoff at $T^\prime_A\approx 6.7 \times 10^4$~s that separates growth as $T^3$ from growth as $T^2$.
Similarly, the averaged periodogram $\overline{S}_{\mathcal{T}}(f)$ portrayed in figure~\ref{bothtemppg}(b) displays a low-frequency cutoff at $f^\prime_S \approx 1.0 \times 10^{-5}$~Hz that separates decay as $1/f^2$ from decay as $1/f^3$, and a high-frequency cutoff at $f_S  \approx 2.5 \times 10^{-2}$~Hz that separates decay as $1/f^3$ from behavior as $1/f^0$.
Thus, the \emph{low-frequency/large-scale} cutoff product $f^\prime_S  T^\prime_A  \approx  0.67$ and the \emph{high-frequency/small-scale} cutoff product $f_S T_A  \approx  0.60$ are comparable, as expected.   
  
The consistency of the scalograms and periodograms confirms that the computations for both measures are sound.

\section{Conclusion and Discussion} \label{sec:concl}

We have presented a collection of periodograms that display the spectral characteristics of ambient temperature fluctuations collected in our laboratory under different conditions. 
A subset of those experiments made use of two independent temperature-measurement systems; the concurrence of the results confirms that our measurement methods are sound. Joint temperature measurements using data collected by both systems attest to the consistency of our results. A corresponding collection of scalograms comport with our periodograms, verifying that our computational methods are trustworthy. It is interesting to note  that the issue of temperature fluctuations has a long history in the annals of physics, dating from the time of Gibbs \cite{gibbs48}, that has not been without controversy  \cite{landau80,kittel88,mandelbrot89,chui92}.

We now proceed to demonstrate that our ambient temperature periodograms match those calculated from data collected at European weather stations over the frequency region where they overlap, providing support for the reliability of our experimental findings. In weather-station temperature analyses, it is customary to model measures such as periodogram in terms of  decreasing piecewise power-law functions \cite{eliazar20}; see, for example, Pelletier \cite{pelletier97}. We show that the $1/f^2$ behavior observed in our periodograms accords with that observed at European weather stations over the same range of frequencies. This comparison is offered as an attestation to the validity of our experimental findings and not as commentary on a particular weather-analysis model. 

\begin{figure}[ht]
\centering
\includegraphics[width=10cm]{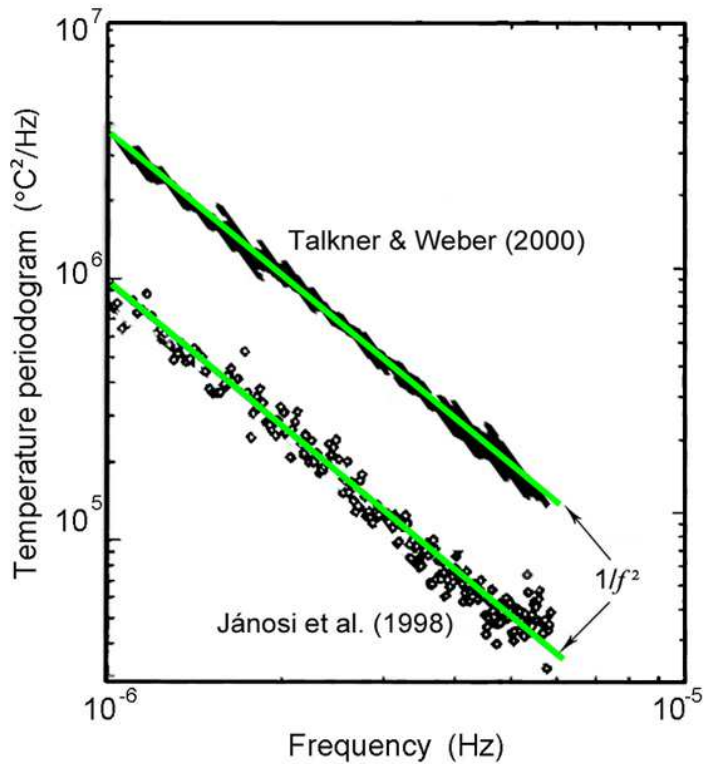}
\caption{Temperature-time-series unnormalized periodograms collected at weather stations in Hungary and Switzerland, plotted on doubly-logarithmic coordinates. The straight lines, which are proportional to $1/f^2$, were chosen to best fit the data. The data are seen to behave as $1/f^2$ over the frequency range 0.1/d $\le f < 0.5/$d, corresponding to $2 < T_f \le 10$~d. In both studies, the maximum frequency reported in the periodogram is $ \frac12 /$d $= 5.79 \times 10^{-6}$ Hz, corresponding to $T_f = 2$~d, as mandated by the rate at which the underlying mean-temperature data were recorded, which was once daily ($T_s = 1$~d). The data in the lower curve are adapted from J{\'a}nosi, Vattay \& Harnos \cite{janosi98} while the data in the upper curve are adapted from Talkner and Weber \cite{talkner00}.}
\label{jan4talk}
\end{figure}

In figure~\ref{jan4talk}, we present temperature time-series periodograms collected at European weather stations in Hungary and Switzerland. The data in the lower curve, adapted from J{\'a}nosi, Vattay \& Harnos \cite{janosi98}, represent the unnormalized periodogram constructed from a temperature time series collected at the Hungarian Meteorological Service weather station in Szombathely (elevation 209~m), using a Hann window. Similar results were obtained for data collected at two other Hungarian weather stations, in Ny{\'i}regyh{\'a}za (elevation 116 m) and in Gy{\H o}r (elevation 108~m). 

The data in the upper curve, which are adapted from Talkner and Weber \cite{talkner00}, represent the unnormalized periodogram constructed from a temperature time series comprising daily mean temperatures collected at the Schweizerische Meteorologische Anstalt weather station in Z{\"u}rich (elevation 556 m), using a Welch window. A nearly identical periodogram was obtained from data collected at the Swiss Alpine weather station on S{\"a}ntis mountain (elevation 2490~m). Temperature data collected at other low- and high-altitude Swiss weather stations also yielded similar periodograms.

The results of our laboratory measurements, presented in figure~\ref{bothtemppg}(b), are in accord with those presented in figure~\ref{jan4talk} over the spectral region that the three sets of data share. In particular, the periodograms of J{\'a}nosi, Vattay \& Harnos \cite{janosi98} and Talkner and Weber \cite{talkner00} exhibit $1/f^2$ behavior over the range $2 < T_f \le 10$~d whereas our averaged indoor-temperature periodogram exhibits $1/f^2$ behavior over the range $1.2 < T_f \le 11.6$~d.

Spectral behavior of the form $1/f^2$ is often ascribed to temperature fluctuations that exhibit a random walk or ordinary Brownian motion. Although one might not expect that the spectral behavior of temperature time series collected in an indoor laboratory would coincide with those collected at a European weather station, the explanation for this conjunction may lie in the work of J{\'a}nosi, Vattay \& Harnos \cite{janosi98}. These authors have compared the spectral properties of temperature measurements conducted in small laboratory convection cells \cite{wu90} with those of extensive meteorological weather stations. The statistical character of the temperature fluctuations in small-scale laboratory experiments may resemble that exhibited by the diurnal atmospheric boundary layer by virtue of the hydrodynamical similarity between these seemingly different systems.

In any case, our temperature periodograms extend to substantially higher frequencies than those of J{\'a}nosi, Vattay \& Harnos \cite{janosi98} and Talkner and Weber \cite{talkner00} because of the far higher rate at which we sampled the temperature, namely every 5~s. And it is at these higher frequencies, in particular over the frequency range $1.0 \times 10^{-5} \le f \le 2.5 \times 10^{-2}$~Hz, corresponding to 40~s $\le T_f \le 1.2$~d, that we observe $1/f^3$ spectral behavior. We have shown that the periodograms and scalograms in our ambient temperature measurements clearly follow such cubic-law behavior; formulating a rationale with respect to why this is so is a work in progress. 

\section*{Acknowledgments}
We are grateful to the Boston University Photonics Center and its director, Thomas Bifano, for extensive financial and logistical support. The Photonics Center provided a dedicated laboratory in which we could control, restrict, and suspend climate control (heating and air conditioning), janitorial services, and the access of personnel. Individual experimental runs, which typically lasted approximately two weeks, were conducted continuously over a period of 18 months, from May 2010 until December 2011. A vast quantity of data was collected during this period; its analysis required the development of specialized software and involved many meetings and extensive correspondence among the authors.

\section*{ORCID iDs} 
Steven B. Lowen https://orcid.org/0000-0002-5243-2548\\
Malvin Carl Teich https://orcid.org/0000-0001-8164-4622\\ 

\section*{References}

\bibliographystyle{unsrt.bst}
%bibliographystyle{iopart-num}
\bibliography{main}

\begin{thebibliography}{10}

\bibitem{hodgkin49}
A.~L. Hodgkin and B.~Katz.
\newblock The effect of temperature on the electrical activity of the giant
  axon of the squid.
\newblock {\em Journal of Physiology (London)}, 109:240--249, 1949.

\bibitem{josefsberg12}
J.~O. Josefsberg and B.~Buckland.
\newblock Vaccine process technology.
\newblock {\em Biotechnology and Bioengineering}, 109(6):1443--1460, 2012.

\bibitem{saleh19}
B.~E.~A. Saleh and M.~C. Teich.
\newblock {\em Fundamentals of Photonics}.
\newblock Wiley, Hoboken, 3rd edition, 2019.

\bibitem{abbott16}
B.~P. Abbott et~al.
\newblock Observation of gravitational waves from a binary black hole merger.
\newblock {\em Physical Review Letters}, 116:061102, 2016.

\bibitem{mohan2020}
N.~Mohan, S.~B. Lowen, and M.~C. Teich.
\newblock Photon-count fluctuations exhibit inverse-square baseband spectral
  behavior that extends to $<1\,\mu${H}z.
\newblock {\em Physical Review Research}, 2:013170, 2020.

\bibitem{mohan10}
N.~Mohan.
\newblock {\em Photon-Counting Optical Coherence-Domain Imaging}.
\newblock PhD thesis, Boston University, Boston, MA, 2010.

\bibitem{oppenheim09}
A.~V. Oppenheim and R.~W. Schafer.
\newblock {\em Discrete-Time Signal Processing}.
\newblock Pearson, New York, 3rd edition, 2009.

\bibitem{lowen05}
S.~B. Lowen and M.~C. Teich.
\newblock {\em Fractal-Based Point Processes}.
\newblock Wiley, Hoboken, 2005.

\bibitem{daubechies92}
I.~Daubechies.
\newblock {\em Ten Lectures on Wavelets}.
\newblock Society for Industrial and Applied Mathematics, Philadelphia, 1992.

\bibitem{haar10}
A.~Haar.
\newblock Zur {T}heorie der orthogonalen {F}unktionensysteme.
\newblock {\em Mathematische Annalen}, 69:331--371, 1910.

\bibitem{graps95}
A.~Graps.
\newblock An introduction to wavelets.
\newblock {\em IEEE Computational Science and Engineering}, 2(2):50--61, 1995.

\bibitem{ryan19}
{\O}.~Ryan.
\newblock {\em Linear Algebra, Signal Processing, and Wavelets--{A} Unified
  Approach}.
\newblock Springer Nature, Cham, Switzerland, 2019.

\bibitem{vetterli95}
M.~Vetterli and J.~Kova{\v c}evi{\'c}.
\newblock {\em Wavelets and Subband Coding}.
\newblock Prentice--Hall, New York, 1995.

\bibitem{teich95}
M.~C. Teich, C.~Heneghan, S.~M. Khanna, {\AA}.~Flock, M.~Ulfendahl, and
  L.~Brundin.
\newblock Investigating routes to chaos in the guinea-pig cochlea using the
  continuous wavelet transform and the short-time {F}ourier transform.
\newblock {\em Annals of Biomedical Engineering}, 23:583--607, 1995.

\bibitem{gibbs48}
J.~W. Gibbs.
\newblock {\em The Collected Works of J.~Willard Gibbs}, volume~2.
\newblock Yale University Press, New Haven, 1948.

\bibitem{landau80}
L.~D. Landau, E.~M. Lifshitz, and L.~P. Pitaevskii.
\newblock Statistical {P}hysics: {P}art {I}.
\newblock In {\em Course in Theoretical Physics}, volume~5, chapter~7.
  Butterworth-Heinemann/Elsevier, Oxford, 3rd edition, 1980.

\bibitem{kittel88}
C.~Kittel.
\newblock Temperature fluctuation: {A}n oxymoron.
\newblock {\em Physics Today}, 41(5):93, 1988.

\bibitem{mandelbrot89}
B.~B. Mandelbrot.
\newblock Temperature fluctuation: {A} well-defined and unavoidable notion.
\newblock {\em Physics Today}, 42(1):71--73, 1989.

\bibitem{chui92}
T.~C.~P. Chui, D.~R. Swanson, M.~J. Adriaans, J.~A. Nissen, and J.~A. Lipa.
\newblock Temperature fluctuations in the canonical ensemble.
\newblock {\em Physical Review Letters}, 69:3005--3008, 1992.

\bibitem{eliazar20}
I.~Eliazar.
\newblock {\em Power Laws: {A} Statistical Trek}.
\newblock Springer Nature, Cham, Switzerland, 2020.

\bibitem{pelletier97}
J.~D. Pelletier.
\newblock Analysis and modeling of the natural variability of climate.
\newblock {\em Journal of Climate}, 10:1331--1342, 1997.

\bibitem{janosi98}
I.~M. J{\'a}nosi, G.~Vattay, and A.~Harnos.
\newblock Turbulent helium gas cell as a new paradigm of daily meteorological
  fluctuations?
\newblock {\em Journal of Statistical Physics}, 93:919--926, 1998.

\bibitem{talkner00}
P.~Talkner and R.~O. Weber.
\newblock Power spectrum and detrended fluctuation analysis: {A}pplication to
  daily temperatures.
\newblock {\em Physical Review E}, 62:150--160, 2000.

\bibitem{wu90}
X.-Z. Wu, L.~Kadanoff, A.~Libchaber, and M.~Sano.
\newblock Frequency power spectrum of temperature fluctuations in free
  convection.
\newblock {\em Physical Review Letters}, 64:2140--2143, 1990.

\end{thebibliography}

\end{document}